\documentstyle{article}

\def\abstract#1{\vskip 7mm 
        \begin{center}{\large Abstract}\par \smallskip
                \begin{minipage}[c]{12cm}
                        \small #1
                \end{minipage}
        \end{center}
}
\def\title#1{\begin{center}{\Large\bf #1}\end{center}}
\def\author#1{\vskip 5mm \begin{center}{#1}\end{center}}
\def\address#1{\begin{center}{\it #1}\end{center}}
\def\et{et al.\ } 
 
\def\ha{H$\alpha$\ }
\def\hb{H$\beta$\ }
\def\lya{Ly$\alpha$\ }
\def\arcs{\ifmmode {''}\else $''$\fi}
\def\mic{~$\mu$m}

\def\ie{{\it i.e.}}
\def\eg{{\it e.g.}}

\def\plotfiddle#1#2#3#4#5#6#7{\centering \leavevmode
\vbox to#2{\rule{0pt}{#2}}
\includegraphics{#1}}
\makeatletter
\@ifundefined{lesssim}{}{}
\@ifundefined{gtrsim}{}{}
\def\vereq#1#2{\lower3pt\vbox{\baselineskip1.5pt \lineskip1.5pt
\ialign{$\m@th#1\hfill##\hfil$\crcr#2\crcr\sim\crcr}}}
\makeatother

\begin{document}

\title{%
  {\large --- Infrared Views of Galaxy Evolution: A Progress Report  ---}
}
\author{Matthew Malkan,\footnote{E-mail:malkan@astro.ucla.edu}}
\address{%
  UCLA Astronomy Division, Los Angeles, CA 90095-1562, USA
}
\abstract{
The two major functions in galaxy evolution that we would like
to measure are the stellar populations in galaxies, and 
their time-derivative, the star formation rate.
Especially at redshifts above 1, both of these measures 
are benefiting greatly from wide-area infrared observations. 
Several space-based and ground-based examples will be discussed.
}

\section{Understanding Galaxy Evolution with Deep Fields}

Since this is a conference talk rather than a journal paper,
I'll start on a philosophical note. Let's reflect on some 
of the good fortune observational cosmologists enjoy.
The task of understanding the contents
of the universe, (let alone their evolution), is so audacious,
that it is remarkable we can even start doing the job.
One reason we can is that gravity helpfully imposes some
clear organization on how matter 
is arranged. %in the universe
%(except the extremes),
Luminous mass appears (at most wavelengths)
to be strikingly organized into
stars and galaxies. Biologists rely heavily on 
their crucial organizing concepts of cell,
organism and species.  As astronomers, we see
the world clearly arranged into self-gravitating planets,
stars and galaxies.
Although these categories span enormous ranges, for
example in luminosity, they obey 
regularities and patterns.  This is why we can
make clear distinctions between what is or
is not a star, and what is or is not a galaxy.
We can, for example, distinguish individuals in
these classes from bound clusters of them.
These are two of the surprisingly easy
categories which define our field.
When we cannot use them, our job gets harder.
%to understand the structure and contents of the universe.
Operationally, the star and galaxy concepts are closely
linked, since we rarely think about one being present
without the other. It is in fact very difficult to learn
much about ``galaxies"  before they began forming many
stars.  I'll discuss one possible exception
to that generalization at the end.

The last half of the 20th century saw dramatic
progress in understanding the evolution and
even the formation of stars.  The first half of the
21st will herald similar advances in understanding
the evolution and formation of galaxies.  These
systems are sufficiently more complicated and
distant than stars that empirical, observation-
driven data must play a relatively larger role than
theory.  (Stellar structure and evolution are 
rarely influenced by external circumstances.
Isolated stars, for example, hardly ever
experience close encounters. Galaxies
often do, and their motions are often dominated by
dark matter we know little about.)

Since galaxies have a wide range of (evolving) properties, 
we need to study large samples of them.
Fortunately, they are so numerous that deep field surveys 
with large array detectors can provide such samples, 
often spanning a wide range of redshift simultaneously.
Two of the most fundamental observables to derive from 
galaxy surveys %of field galaxies
are 1) their luminosity function (a measure of their
population of formed stars) and 2) their rates of
recent and ongoing star formation.
In deep fields we can study how both of these evolve through
cosmic time.

Of the results obtained from the Hubble Deep Field (HDF),
the most far-reaching is that the  star formation
rate is high at $z \sim 3$ (\cite{ma96},\cite{ma98}),
with a possible %modest 
decline at $z \sim 4$.
When combined with
low-redshift data (e.g. \cite{lil}), there is evidence
for a peak in the comoving SFR at $z\approx 1--2$,
around a third of the Hubble Time.   The apparent resemblance between
this star formation rate curve and the evolution of quasars
has been noted.

%next we can consider morphology, interactions
%I will argue that 
For both of the main goals in
the study of galaxy evolution--measuring
the stellar population, and its first time derivative--observations with 
large-format (\eg million-pixel) infrared detectors are 
essential. % for measuring the evolution of both of these.
I will illustrate this by reviewing several applications
of infrared observations.
Some of these were also discussed at a Tokyo conference of
the Birth and Evolution of the Universe (\cite{mat01}).
Even though only two years have passed, so many advances
have been made, that an update is due.

\section{Detecting Evolving Stellar Populations Across 
Cosmic Time } %/Galaxies}
%\subsection{  Bright Ages" Relatively Unexplored}

Deep multicolor imaging is our most powerful tool
for understanding galaxy evolution. 
Broadband filter measurements can go very deep; the
resulting spectral energy distributions
provide estimates of the photometric redshifts
of distant galaxies.
The Hubble Space Telescope, with its superb spatial resolution, 
provides deep detections, along with detailed measurements of galaxy morphologies,
in small fields.
%this, but it has covered only a
%but it has proven too expensive
%to pursue in more than a tiny 
%handful of small fields.
Large CCD cameras on ground-based telescopes provide
complementary coverage of much larger areas
(e.g. in \cite{ouch}, and
the Sloan Digital Sky Survey.)
The next step, 
%The great observational efforts 
%including the 
intensive spectroscopic follow-up
(\eg, \cite{st96})
has been rewarded with 
the detection of many hundreds of galaxies
at redshifts of around 1 and 3.
Most galaxies at $z \ge 3$ have been discovered using the powerful
Lyman break/U-band drop out selection method (\eg, \cite{st99}).
%Adelberger 2001).  astro-ph/0101144).  
%A faster technique is the search for Lyman limit 
%breaks, in ``UV-dropout" objects for example . Unfortunately, 
This method, which is currently confined to z 
$\sim$ 3 - 4,  does not of itself provide an accurate confirmed 
redshift.  Follow-up optical spectroscopy is still needed. 
%Unfortunately, 
The Lyman break method falters at $z<2.7$ because
at that redshift the Lyman limit has not yet cut out much of the
continuum in the U-filter. %  employed;
%While clever use of special filters can help,
Lyman break galaxy (LBG) surveys are
not possible in the redshift 2 range.  

The intermediate redshift range around $z \sim 2$ appears to be
%we are specially targeting--around z=2--is around
the epoch during which the familiar components of modern galaxies assembled.
%decrease in the number density of damped $L\alpha$ absorbers
%and the QSO $L\alpha$ forest are consistent with an overall
%tendency for gas to be turned into stars (Pei \& Fall, 1995, ApJ,
%454, 69).
%Present results indicate that luminosity density and
%global star formation rates increase with redshift out past z=1 and
%are falling or have levelled off when measured around $z\approx 3$.
A significant percentage of the universe's
evolution ($\approx 3$ Gyr), 
or 25\% of its total age, occurred from z=1 to 3, 
as opposed to the more distant and better-studied z=3-4 range, 
which covers $\approx 0.5$ Gyr, or only 5\% of cosmic time.
This intermediate epoch can be dubbed the ``Bright Ages", since it
appears to be the time when most stars
in the Universe formed (and most heavy elements were produced).
%Even though the epoch
%corresponding to $z\approx 2$ is the most energetic
%period in a typical galaxy's history and consequently
%the point in time where the
%majority of its stars are created, remarkably few of these galaxies have
%ever been observed and spectroscopically confirmed.

However, our observational knowledge of galaxies in the crucial range of $z=
1.5-2.5$ is surprisingly sketchy.
Due to severe limitations of optical search and confirmation methods, we are in the odd
position of knowing more about the most distant galaxies than those closer.
%especially in the relatively unexplored but crucial range
%$1 \le z \le 2.7 $
Continuum observations at $\lambda \ge 4000 \AA$ are needed to
assess the build-up of evolved stars. Similarly, the most useful
emission
lines for measuring gas photoionized by
recently formed stars are in the rest-frame optical.
Thus the most reliable measures of total stellar mass as well
as current star formation rates (the first derivative of the former)
are obtained in the near-infrared for redshifts in the Bright Ages.

\subsection{Near-Infrared Searches for Galaxies in the Bright Ages}

Some of the early successes in identifying non-active,
non-lensed galaxies at 
high redshifts %in general 
came from searches for strong emission lines with narrow-band filters.   
This method provides accurate redshifts, but selects from a limited range of  
redshifts and has yielded only a modest number of detections (\eg, \cite{mtm95},
\cite{mtm96}, \cite{fran}). 
%McLean 1995; 1996; Pascarelle {\it et al.} 1996, and Francis {\it et 
%al.} 1996). 

%developments now allow us to overcome these limitations:
%combined with deep near-IR photometry
%that extends to 
Optical multicolor photometry alone works well at measuring
the Balmer break up to $z \sim 1$.
%discuss adelburger optical balmer breaks up to z=1.xx
Infrared photometry makes it possible for us to identify
Balmer break galaxies at $1.5 \le z \le 2.5$.
New large-format detectors are able to make
sensitive surveys in the near-IR wavebands, especially
with the key 2.2\mic\ (K) band.
This allows us to measure the
second strongest spectral feature in galaxies--the Balmer break,
from around 4000\AA\ to 3650\AA.
This blue-side drop is nearly always strong in populations of
young or old stars (although its wavelength shifts slightly
redward in older galaxies).

\begin{figure}
\plotfiddle{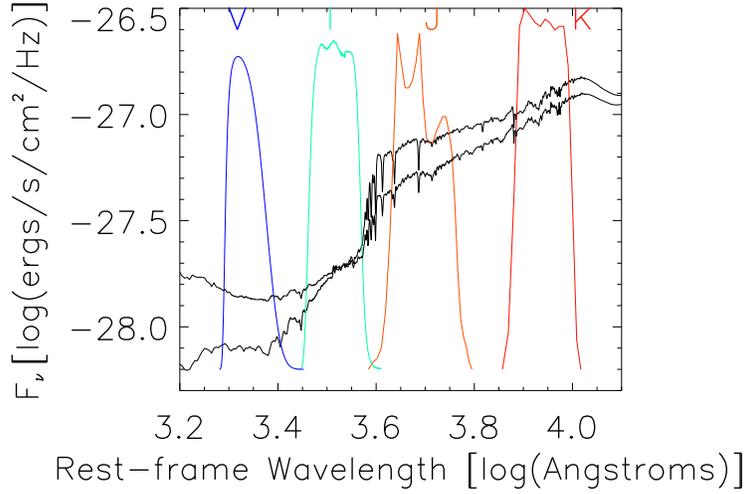}{1.3in}{0}{60}{60}{-180}{-10}
\caption{ The VIJK filter set superposed on typical galaxy spectra at
z=1.5.  The representative Spectral Energy Distribution for the galaxy
is from the GISSEL spectral synthesis models.  The Balmer break produces
a very red I-J color. As the redshift increases above 1.5, the shift in the Balmer
break makes J-K redder than 2.
}
\end{figure}

Figure 1 shows the V, I, J and K bandpasses superposed on the model
spectrum of a galaxy at z=1.5.
To reach z=2.5 requires good photometry in the
K filter, so that at least one point in the spectral energy distribution
is cleanly on the {\it red} side of the Balmer break. 
Two-color %(e.g., (V-I)/(J-K)) %in addition to our
%straight photometric redshifts,
plots are useful for identifying galaxies likely to lie in the Bright Ages epoch.
An example is shown in Figure 2. 
Models with a variety of star-formation histories are shown for
galaxies with $0 \le z \le 1$ by the crosses; models with $ 1 \le z \le 2$ by the
diamonds, and $2 \le z \le 3$ with the small dots.
In the astro-ph version I color-coded these model predictions based on their 
V-I colors.  The open squares show galaxy photometry from one of our
deep fields imaged on the Lick 3-m telescope.

\begin{figure}
\plotfiddle{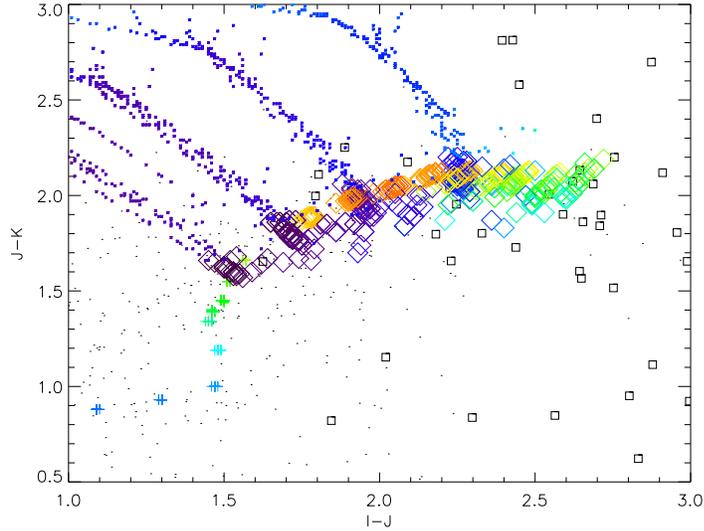}{0.5in}{0}{60}{60}{-180}{-10}
\caption{ The the I-J,J-K plane for model galaxies at redshifts from 0 to 5.
In the astro-ph version, additional V-I
information is encoded in color (with the bluest values shown as the bluest
points).  
%The models plotted sample the
%z$>1$~galaxy colors -- the redshift range of highest interest to this
The colored points outline tracks for Bruzual and Charlot
stellar evolution models of galaxies with initial formation
epochs at high redshift which then have superposed more recent
starbursts of varying strengths. The limited grid of starburst epochs
(z=2.6, 2.8. 3.0, 3.5, 5) produces the several different locations of
the ``peeling back' in the two-color diagram. Note that the rest-frame
UV light (shifted into the B$+$V bands) is extremely sensitive to the
temporary presence of young stars, while the rest-frame optical (in
the J and K bands) is insensitive to them.  The open squares show
our measurements of galaxies in one of the deep fields we observed with
the Gemini 2-channel infrared camera on the Lick 3-m telescope.
}
\end{figure}
 
The 3600/4000\AA\ break is
less strong than the Lyman break. Nonetheless, by fitting the full 
galaxy multiwavelength spectral energy distribution to
redshifted galaxy templates we can estimate its photometric
redshift to an accuracy of one or two tenths in z
(\eg, \cite{koo}, \cite{fern}, \cite{brun}, \cite{hogg}).
%1999, ApJ, 513, 34; Hogg et al. 1998, ApJ, 504, 622).  
Infrared flux points greatly
improve the accuracy of the method \cite{con97}. 
To demonstrate the power of this method, 
we present a six band (B through K) picture of 
an very red object %(ERO) 
%taken from the data in Fig. 3
\cite{com02}. Figure 4 shows the 
photometric redshift fit to its BVIJHK spectral energy distribution, using
%Figure 4 shows an example from deep multicolor BVIJHK imaging,
%of a galaxy with a photometric redshift (from 
Hyperz (\cite{bol}). %Bolzonella et al. 
%2000, A$\&$A, 363, 476))
%(Colbert and Malkan 2001).
%see also Brunner \et 
The infrared bands  were crucial not only 
in identifying this as an interesting object,
but allowing a fit to data beyond the 4000\AA\ break. 
The fit estimates its redshift,  z=1.23 +-0.25,
and also that it is a 0.5 Gyr old burst of star formation with $A_V = 1.2$ mag 
of dust extinction. The infrared photometry provides a fairly good
estimate of the total stellar mass.

A lot of what we think we know about high-redshift galaxies
rests on the limited statistics of photometric redshifts in the
Hubble Deep Fields, mostly HDF-N.
%The HDF-N field has about 80 , with estimated
%redshifts between 2.6 and 3.8, and a somewhat smaller
%number of  with inferred redshifts above 3.8.
%Only about a dozen of the ``U-dropouts" (from the Lyman break
%at $z \ge 2.6$) in HDF-N and
%four of the
%B-dropouts (from the Lyman break
%at $z \sim 4$) have actual spectroscopic confirmations.
The HDF-N has 150 galaxies with photometric redshifts
in the Bright Ages ($1.8\le z \le 2.6$).
However, only 8
of these are spectroscopically confirmed after extensive efforts,
and {\it no} Bright Ages galaxy candidates in HDF-S 
have confirming spectra yet.
Major conclusions have been drawn from analysis of one
$2.6^{'} \times 2.6^{'}$ image of the Universe.
This is particularly dangerous since LBG surveys have found large
field-to-field variations--up to a factor of four in surface
density--due to large-scale structures, which also appear as
spikes in redshift histograms.  Since our view of galaxies
depends strongly on which sight-line we observe,
{\it it is imperative to investigate  galaxy evolution %the star formation history 
in several regions with similarly deep data}.
Many efforts, both ground- and space-based, are currently
underway to do this, for example at Subaru (\cite{mai}), VLT (VIRMOS/DEEP)
and NOAO (Deep Wide-Field Survey).

\begin{figure}
\plotfiddle{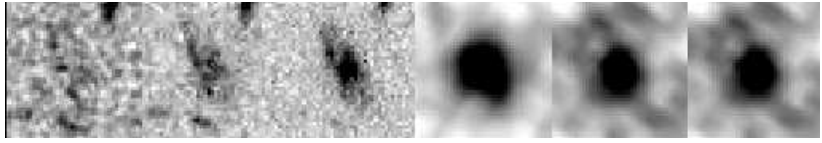}{.5in}{0}{60}{60}{-180}{-200}
\caption{Mosaic of reduced multicolor images of a 4 x 4 \arcs\ box around
a galaxy in the 53W002 field with photometric redshift $z=1.12$.
Wavelength increases from B at the left to K on the right; the object
can qualify as an ``Extremely Red Object" (ERO), with $V-K>6$.
The smooth round bulge becomes progressively more dominant at longer wavelengths,
while at the shorter wavelengths the galaxy appears patchy with irregular arms.  
%At this redshift the rest wavelength
%of the K-band image is approximately at 1 $\rm \mu m$.
}
\end{figure}

\begin{figure}
\plotfiddle{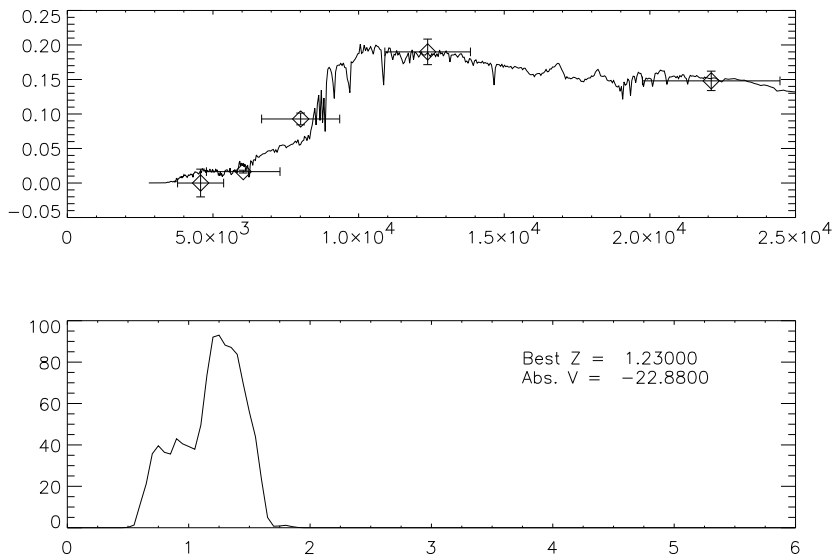}{2.7in}{0}{65}{65}{-210}{-260}
\caption{BVIJK photometry we measured for an ``extremely red" 
galaxy in the MTM0953+549 field. Note the sudden jump at
the Balmer break, which makes a very large V-J color,
even though  the J-K is not unusually red. 
The best-fitting spectral template, at z=1.23 is overplotted
in the top panel.  
%It includes a 0.5 Gyr burst of star formation
%reddened by $A_V = 1.2$ mag.  
The lower panel shows the probability 
distribution of
acceptable redshifts fitted by the photometric redshift engine.
The best fit implies the galaxy has $M_V = -22.9$. 
}
\end{figure}

\section{Evolution of Star Formation is also well traced in the infrared}

A region's ``current" star formation rate (SFR)
is approximately proportional to its population of upper main
sequence stars, because of their cosmically brief lifetimes.
The presence of massive stars is manifested in several observables
which are not strongly produced by old stellar populations:  
a) direct photospheric radiation, predominantly in the UV.  
At lower frequencies it extends into the optical,
where it produces blue colors in a stellar population.
The high-frequency tail extends
above 1 Ryd, and includes a substantial fraction of ionizing photons,
in the hottest stars.  Much of this radiation is absorbed by
gas and re-emitted as  b) nebular emission lines, mostly
in the optical and infrared.  The most
massive stars hardly live long enough to completely leave the 
dusty environments in which they were born.  Thus, much of their
remaining continuum emission may be absorbed by interstellar dust
grains, which are warmed to c) re-emit the power thermally in the infrared.
Finally, massive stars are the most violent sources of mechanical energy in 
the ISM, 
driving winds and supernovae.  A fraction of this power accelerates electrons
which then emit  d) synchrotron radiation.

After a great deal of work on the alternatives, it is \ha line emission which remains
the ``gold standard" of SFR measurements. It is equivalent
to counting the ionizing continuum photons from the hottest
(youngest) main-sequence stars at a wavelength that is relatively 
insensitive to absorption.  
%Ideally, 
The observable line luminosity is supposed to
equal a constant times
the aggregate SFR: $L(H\alpha) = 1.1 \times 10^{41}$ erg/sec per $M_{\odot} / year$
\cite{ken}. 

By contrast, the other leading SFR indicator 
at high redshift--the ultraviolet continuum--is
far more sensitive to dust reddening. 
The integrated star-forming luminosity  of
those LBGs {\it detected} 
needs to be corrected upward for extinction by a factor of 
2 -- 7  (\cite{pet98}, \cite{pet}, %Cohen et al. 1998;
\cite{adel}, \cite{hop}, and \cite{tep00}).
Furthermore, these LBGs were selected to be the {\it least} dusty ones. 

We can obtain a better idea of why the line and continuum SFR estimates
disagree (methods a and b above) by looking
closely (with HST) at nearby star-forming galaxies.
The contours in Figure 5
show our HST imaging of the (continuum-subtracted) \ha emission in the 
small star-forming galaxy NGC 2328.  The greyscale shows the optical 
continuum, which has a considerably different spatial distribution.  
Note the numerous contour peaks with little or no associated optical light.
These HII region complexes with significant reddening account for the
majority of the current star formation in this galaxy.
Another even more dramatic example of a dust-enshrouded starburst is
seen in the nearby galaxy NGC 5253 \cite{jean}.

\begin{figure}
\plotfiddle{2328HC.ps}{1.5in}{0}{20}{20}{-60}{0}
\caption{Our WFPC-2 continuum-subtracted H$\alpha$
image of ``Early-Type" Galaxy NGC 2328 from Glassman and Malkan \cite{gm},
shown as a greyscale plot. The optical continuum is shown with overplotted
contours. The field of view is roughly 6 x 10\arcs.
}
\end{figure}

Although \ha  gives a more complete census of recent star formation,
it cannot be measured with
CCD spectrographs for $z \ge 0.5$.  Even the weaker and less reliable
\hb line is shifted out of the optical range at $z \ge 1$.
Fortunately, they are still measurable in the near-infrared.
Two methods--one ground-based and one space-based--have been
successful.

\subsection{Narrow-Band Infrared Imaging of Emission-Line Galaxies}

Large-format multislit spectrographs are not yet available  in
the near-IR.  This will change when 
instruments such as the NIRMOS (Near-infrared
Multi-Object Spectrograph) begin working on the VLT,
and Flamingos (in multi-slit mode) on Gemini. 
For now, ground-based searches are restricted to either narrow 
spatial or spectral windows.
% single slit spectra of 1 or 2 high-z candidates at a time and
%tens of square arcsecs of blank sky
The latter case is obtained 
with narrow-band interference filter imaging.
To speed up the detections, there is often a
pre-selection of special fields. These are supposed  
to have excess galaxies at the targeted redshift, because
some object in the field is already known to have that z.
Several dozen galaxies at redshifts above 2 have been
discovered by detecting their \ha line emission
(\cite{mtm95}, \cite{mtm96}, \cite{bunk},\cite{kaw}, \cite{mann}).
More discoveries will be coming soon with the availability
of 1K x 1K near-infrared array detectors.

\subsection{Grism Spectroscopy } % of Emission-Line Galaxies}

Slitless spectroscopy receives the full sky
background from the entire spectral window at each point on
the detector. It is therefore impractical with ground-based
infrared telescopes.  However, it has been demonstrated to
work very well in space, where the backgrounds are much lower.
In particular, the NICMOS Camera 3 on HST has an objective
grism that disperses the spectra of all objects within its
50\arcs\ field-of-view.
Since it was able to operate simultaneously with other HST
instruments, it obtained a substantial number of deep
parallel pointings on random fields.
The largest set of these were reduced and analyzed by
McCarthy \et \cite{pmc}. The exposures of a few orbits had
a 1 in 3 chance of detecting a line-emitting object in
the field; in the longer parallel observations, these
odds increase to nearly 100\%.
In addition,
a single very deep NICMOS grism field was observed during
a 3-day pointing in the Continuous Viewing Zone
(\cite{mat02}).

The standard assumption in deep near-IR grism surveys of random
fields is that any single emission line detected is 
redshifted \ha. In only one of the dozen cases tested so
far has this assumption been shown to be incorrect; it is
likely true better than 90\% of the time.
This allowed Yan \et \cite{yan} to estimate the global luminosity
function of recent star formation in the $0.8 \le z \le 1.7$ universe.

\begin{figure}
\plotfiddle{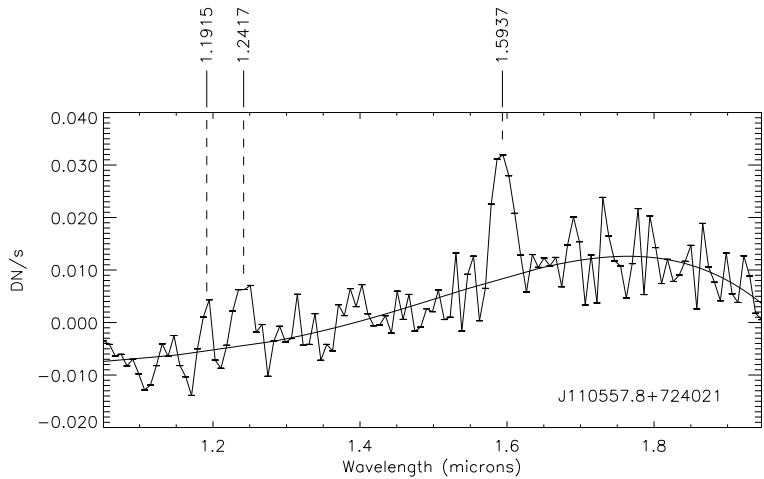}{1.4in}{0}{60}{60}{-200}{-230}
\caption{Extracted spectrum of a star-forming galaxy
discovered in a deep NICMOS-grism parallel observation
(Malkan \et \cite{mat02}). At a redshift of 1.49, this low-resolution
spectrum shows both
\ha and \hb, as well as a broad blend of the two [OIII] doublet
lines, the redshifted 5007 and 4959\AA\ emission.
}
\end{figure}

\begin{figure}
\plotfiddle{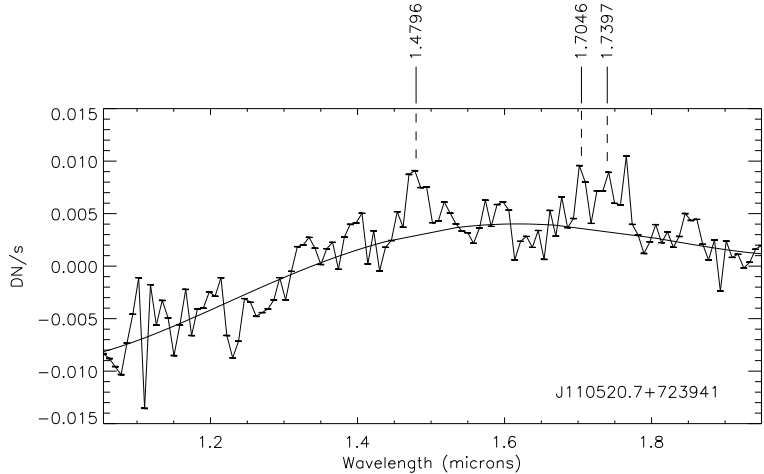}{2.2in}{0}{60}{60}{-200}{-225}
\caption{Another star-forming galaxy in the same NICMOS
parallel field \cite{mat02}. This is the only grism spectrum
in which none of the detected emission lines is \ha.
The lower redshift, $z_e = 0.35$, allows the
detection of the He I 1.083/$P\gamma$ 1.093\mic\ blend and
the [FeII]1.257 and $P\beta$ 1.281\mic\ emission lines.
}
\end{figure}

In some cases, the difficult confirmation with optical spectroscopy
is not even necessary because a second emission line is also
detectable simultaneously in the NICMOS grism image.
Figures 6 and 7 show deep grism spectra (observed wavelengths) 
in which emission lines
other than \ha are detected.
In Figure 6 the additional lines on the blue side are 
$H\beta$, and the [OIII]5007/4959 doublet, at $z_e=1.49$.
In Figure 7 the only  emission-line spectrum
which does {\it not} show \ha is shown.
This foreground  ($z_e=0.35$) galaxy shows
the strong emission lines of He I 1.083/$P\gamma$ 1.093\mic,
[FeII]1.257 and $P\beta$1.281\mic\ \cite{mat02}.
 
In its first incarnation, this camera was defocused by
the distortion of the NICMOS dewar.
This degraded the spatial resolution to about 0.4\arcs.
The resulting spectral resolution was only $R \sim 50$.
Hopkins \et \cite{hop} used NICMOS Camera 3 for pointed observations
of suspected high-redshift galaxies during a campaign when
it was in focus.
Both their observations and the parallel ones
indicated that SFR rates estimated from \ha emission
in galaxies at $ 0.8 \le z \le 1.7$ are several times
higher than those estimated from their UV continuum.
 
\begin{figure}
\plotfiddle{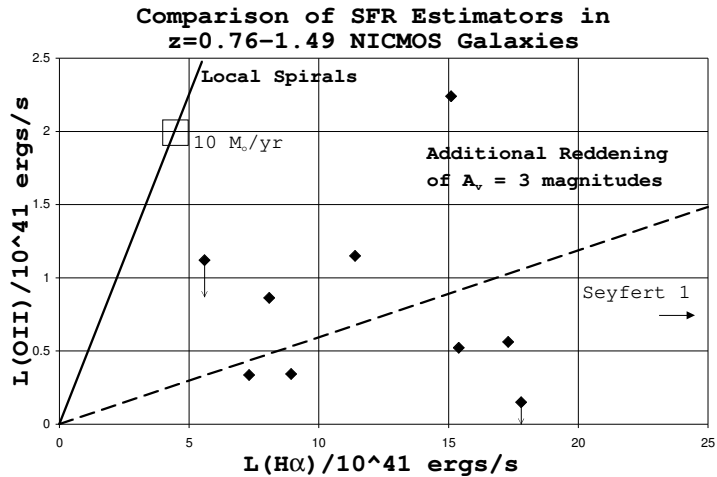}{1.9in}{0}{40}{40}{-180}{-20}
\caption{Comparison of star formation rates estimated from
[OII]3727 and \ha line emission, in $z \ge 1$ galaxies
discovered by the NICMOS grism. Typical uncertainties on the points are 20--30\%.
The open square indicates the line luminosities predicted by
the local relations for an SFR of $10 M_{\odot}/year$.
All of these 9 galaxies observed at
Keck by Hicks \et \cite{hicks} have relatively weaker [OII]
emission than would be predicted by their \ha line, based
on local spiral galaxies (which are described by the solid line).
Instead, their [OII]/\ha ratios are suggestive of additional
internal dust absorption, which might be as large
as $A_V = 3$ mag (shown by the dashed line). }
\end{figure}
 
These large numbers of strong \ha\--emitting
galaxies already indicate that search methods using shorter
wavlengths are incomplete. Similar incompleteness in
surveys of star formation rates has
also been found in the local universe by Sakai \et \cite{ss}.
If short-wavelength surveys really do miss a substantial
amount of obscured star formation activity, we should
examine galaxies
that are found in the near-IR, to see if they are
indeed dustier than those discovered by optical surveys.
In Figure 8 we compare the two emission-line indicators of current
star formation rates--\ha and [OII]3727 luminosities.
The nine data points represent the NICMOS-selected 
galaxies, with followup Keck optical spectra from Hicks \et \cite{hicks}.
The solid line shows the observed average ratio in local spirals,
which does not fit the NICMOS galaxies.
In these IR-selected galaxies the [OII] line systematically
underestimates the SFR.  
This discrepancy between the two indicators cannot
be predicted from the broadband colors, which are not often
extremely red.  
Many of these galaxies
must contain additional internal reddening around their
star-forming regions. The dashed line shows an
indicative amount, of $A_V = 3$ mag.
If the ten galaxies we measured are representative of \ha selection
in general, it produces samples of galaxies which a) would mostly
be missed by optical search methods; and b) still need a significant
upward correction for dust extinction.  %, perhaps by as much as a factor.
We need  \ha and short-wavelength SFR measurements for more
of the {\it same} galaxies to see how much overlap there is
between those selected with optical and IR searches.
%and the dashed line shows 3 magnitudes of additional internal extinction.
%even \ha-selection requires reddening correction

\subsection{Nebular Diagnostic Line Emission at Higher Redshift}
\ha is redshifted beyond the prime sensitivity of
infrared detectors at $z \ge 2.6$. By unlucky coincidence, the rising
thermal background makes it impossible to compare star formation
rates from \ha with those estimated from the ultraviolet continuum
in the LBGs.
Thus the only way to make a {\it direct} comparison of SFR indicators
(\ie, on the same galaxies) is by replacing \ha with another 
emission line that is bluer in the optical rest-frame, and less reliable.
%could be considered a .
The advantage of using \hb as this ``silver standard" SFR estimator
is that its intrinsic strength relative to
\ha is known accurately from recombination theory \cite{pet}. The [OIII] 5007 and
4959\AA\ doublet, however, is usually two or three times stronger, and is not so affected
by underlying stellar absorption. 
In Figure 9 we compare the star formation rates predicted by
[OIII] and UV continuum emission in eleven LBG's at $z \sim 3$.
A few galaxies lie near the bottom of the graph, consistent with 1-to-1 
agreement.  However in most of them, the [OIII] line emission predicts
a higher SFR by a factor of 3 to 7 times.  This is again consistent with
the amounts of internal dust reddening that have been inferred from
other observations.

\begin{figure}
\plotfiddle{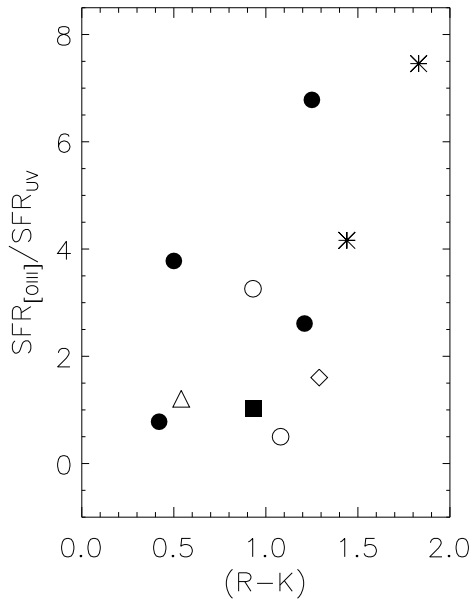}{2.0in}{0}{50}{50}{-160}{-15}
\caption{ The ratio of star formation rates estimated by [OIII]5007
and UV continuum emission from 10 Lyman break galaxies
\cite{tep00}.  [OIII] for the open circles and asterisks 
was measured from
narrow-band imaging; for the remaining points it comes from 
slit spectroscopy. The [OIII] line reveals several
(up to seven) times more current star formation than does
the UV continuum, especially in the redder galaxies.
This is probably largely attributable to internal dust reddening.
}
\end{figure}

%Future Prospects
\section{The Promise of Space Imaging Surveys in the Thermal Infrared}

The thermal dust emission,
(method ``c" above)  can estimate the
high-redshift SFR with hardly any sensitivity to extinction.
Instead of measuring the primary signatures of hot young stars--
their UV continuum and resulting ionized gas--we can measure
this emission after it is reprocessed by dust, since that is where 
much (probably most) of the energy ends up before escaping the galaxy.  
This secondary thermal
continuum is emitted mostly at 25--60\mic.  Since this continuum emission can
dominate the entire bolometric luminosity of a dusty star-forming
galaxy, it is now detectable out to high redshifts.

First, it can be detected globally, by wide-area measurements
of the Diffuse Infrared Background (DIRB) radiation.
New observations continue to define the DIRB more and more accurately,
over a wide wavelength range (from 2--500\mic).
It is surprisingly bright, and requires strong cosmic
evolution in the population of dusty galaxies at $z \sim 1$
(\eg, \cite{ms98}).
``Backwards evolution" models, which are based on our empirical
knowledge of infrared emission from local galaxies, have also
improved.  Figure 10 illustrates a recent comparison between
these models and DIRB observations, from Malkan and Stecker \cite{ms01}.
Two plausible evolutionary scenarios are shown -- a baseline
model with pure luminosity evolution going as $ (1+z)^3$,
and a ``Fast" evolution scenario that reaches the same maximum
luminosities at an earlier z, 1.4 rather than 2.0.
As shown in Figure 11, this latter scenario is somewhat
more consistent with independent estimates of the cosmic
SFR evolution, compiled by Rowan-Robinson \cite{mrr}.

\begin{figure}
\plotfiddle{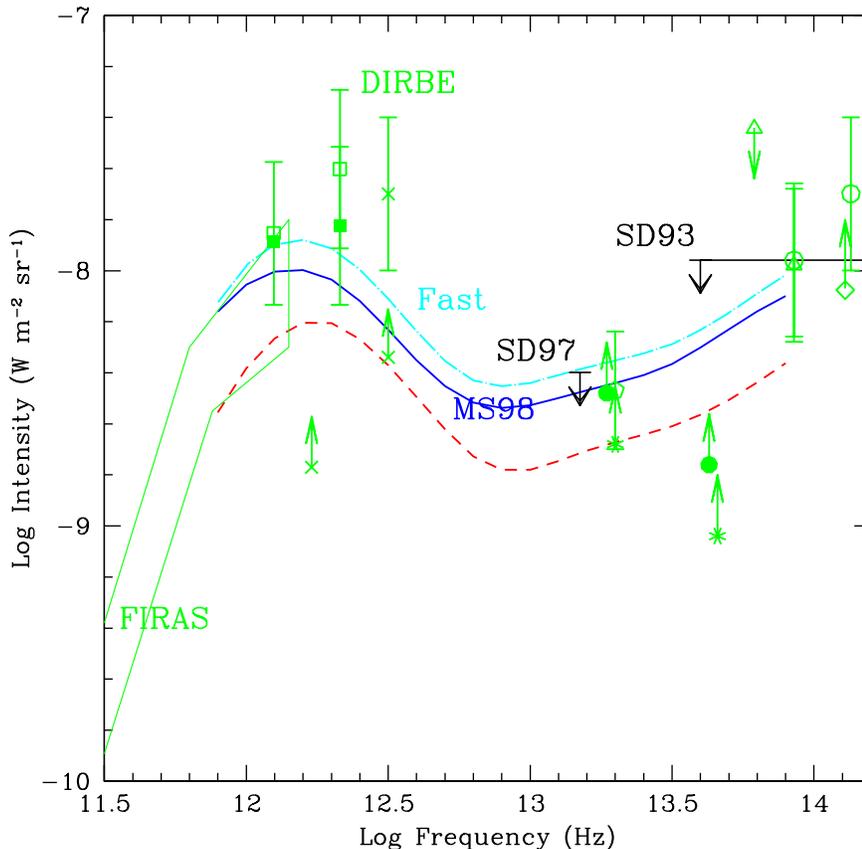}{3in}{0}{60}{60}{-180}{-110}
\caption{Model predictions of the Diffuse Infrared Background
Emission from galaxies, compared with observations, from
Malkan and Stecker \cite{ms01}. The ``best guess"  baseline
evolutionary model shown
by the middle line, assumes that infrared luminosities of
galaxies increase as $(1+z)^3$ from z=0 back to z=2.
The most rapid evolution model, shown by the upper (``Fast")
line, assumes luminosity evolution as $(1+z)^4$ back to z=1.4,
and is also marginally consistent with all the observations.
The lower limits are set by deep number counts obtained with
the Infrared Space Observatory. The two lines with upper limits
are set by the absence of high-energy absorption in the 
gamma-ray spectra of distant blazars, from Stecker and DeJong.
Most of these strong observational constraints became
available {\it after} the MS models were first published.
}
\end{figure}

\begin{figure}
\plotfiddle{mrrfig.ps}{2.5in}{-90}{50}{50}{-200}{290}
\caption{Comparison of the luminosity evolution model
from Malkan and Stecker (1998 and 2001, \cite{ms98}, \cite{ms01}) with estimates
of the global evolution of star formation rates. 
The baseline evolution, and
rapid evolution models of MS are shown by the light and
dark solid lines, respectively.
Both assume no evolution from z of 2 up to 5.  Error bars
are estimates  from
Rowan-Robinson \cite{mrr} based on a variety of observational methods,
with substantial corrections for dust extinction applied.}
\end{figure}

Upcoming deep IR surveys from space will
soon resolve this background. 
The key is surveying enough area with sufficient sensitivity
to detect ``typical" galaxies at the desired redshifts.
``Typical" can be defined by the characteristic knee in
the galaxy luminosity function, which is well defined
at z=0, and appears to be more luminous at high
redshifts \cite{st99}.
Although such ``$L_*$" galaxies are not very numerous
in a standard Schechter luminosity function, galaxies
of $L_*$ or brighter do produce more than 16\% of the
total integrated galactic luminosity.  And in flux-limited
samples, where the number of galaxies detected gets weighted
like $L^{3/2}$, half of the total luminosity will come from
galaxies of 0.7 $L_*$ and brighter.
Figure 12 shows the Malkan and Stecker predictions for 
the observed fluxes of $L_*$ galaxies back to redshift 5,
for 3 different wavelengths. Note that they do not fade
much from z=0.5 to 2 because of their strong luminosity
evolution.
%The key question is sensitivity: can deep surveys 
%Lstar detectability graph

\begin{figure}
\plotfiddle{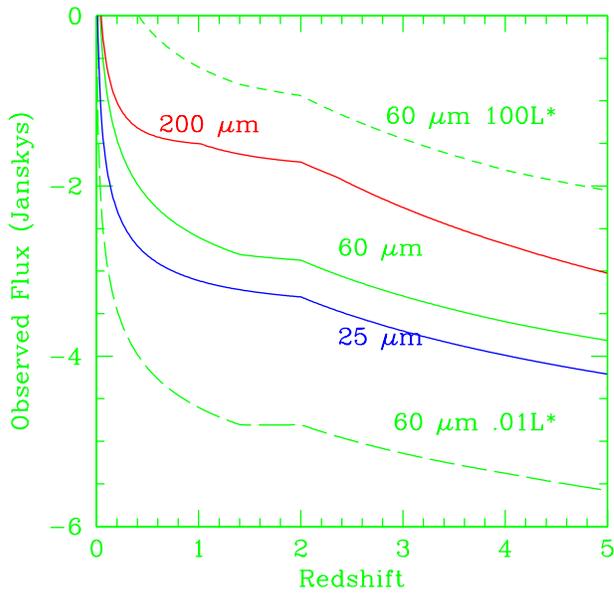}{2.0in}{0}{50}{50}{-180}{-120}
\caption{Predicted brightness at infrared wavelengths of
an $L_*$ galaxy, from Malkan and Stecker
\cite{ms01}, at 25, 60 and 200\mic.  The dotted lines show
the predicted brightnesses for galaxies 100 times above
and below the characteristic $L_*$ knee of the luminosity function.
There is a kink at z=2, beyond which the galaxy luminosity evolution 
is assumed to level off.}
\end{figure}

SIRTF, in its ``First Look Survey", and Astro-E (IRIS), in a wider sky survey,
will have the sensitivity needed to detect
the typical $L_*$ galaxies  %much brighter 
at $z \ge 1$ which are expected to dominate the DIRB.
As shown in Figure 13,
the longest cosmological reach is provided by observations
at 6\mic\ (with SIRTF) and 200\mic\ (with FIRST), since these wavelengths are redward of
the two peaks in the spectra of galaxies. This results
in ``negative K-corrections", the reverse of the usual situation
in which higher redshifts shift a relatively fainter part of
the galactic spectrum into our fixed observing bandpass.

\begin{figure}
\plotfiddle{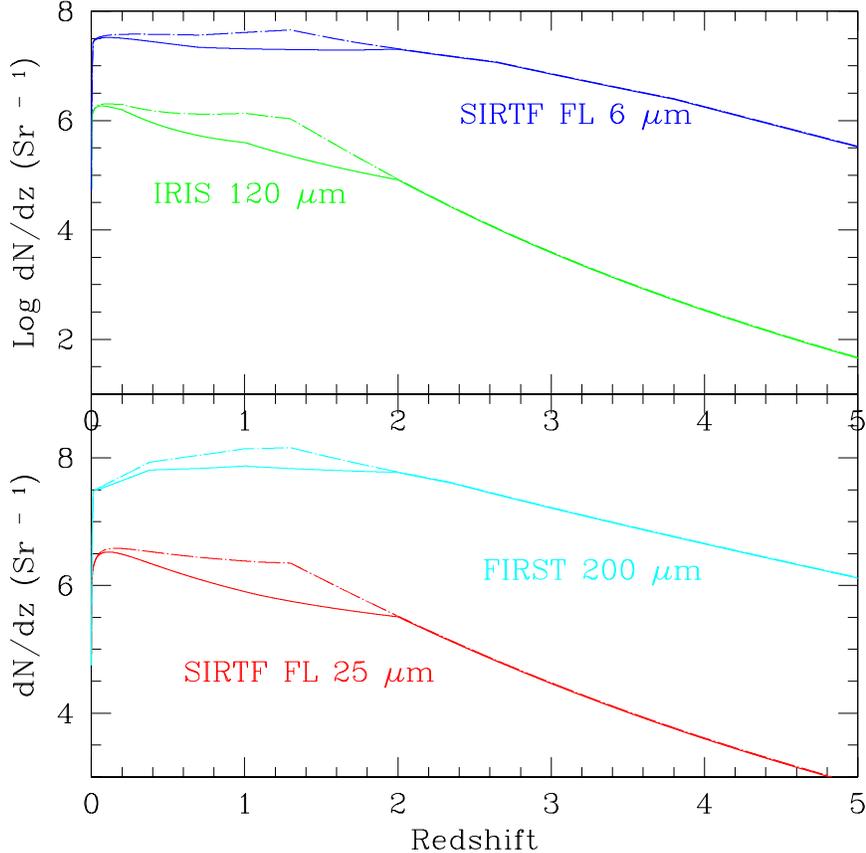}{3in}{0}{60}{60}{-180}{-100}
\caption{Predicted logarithmic number counts per unit redshift for
galaxy luminosity evolution models in Malkan and Stecker \cite{ms01}.
The higher line, shown by dashes, illustrates the ``Fast evolution"
model.  }
\end{figure}

\section{Damped \lya Absorption from Pregalactic Objects?}

Implicit in the previous discussion is the assumption that the ``birth
of a galaxy" is operationally equivalent to the epoch of its first major
burst of star formation.  This is usually the first chance we have
to detect objects which will eventually evolve into modern galaxies.
However, damped \lya absorption systems (DLAs), detected as neutral hydrogen  
troughs ($N_{HI} \sim 10^{20 - 23} cm^{-2}$) in the spectra of  
high-z quasars, % independent detection technique.   
detect baryonic concentrations independent of  star formation  
or nonstellar nuclear activity.  
DLAs contain most of the H I in the universe.
%Whether they are the large disk progenitors of  
%todays spiral galaxies, proto-galactic clumps, dwarf galaxies, compact objects,  
%or low surface brightness galaxies %, or fragments lit up by nearby quasars 
%-- is not yet settled.

%\subsection{The Limitations of Previous Ground- and Space-Based DLA Imaging}
 
%\noindent {\bf  The Need for Space-Based Imaging of DLAs.}  
Imaging searches
for galaxies associated with DLAs 
are not new, but they have been almost impossible to 
carry out from the ground \cite{lowen}.
%{\it et al.} 1995).
They are best 
performed in the near-IR, where the DLA starlight should be relatively bright 
compared to the background quasar.
Unfortunately,  
a large enough H I column density to be detected as 
a DLA is only expected to occur at small 
impact parameters through a galaxy (or protogalaxy) -- within 10 or 20 kpc 
of its center (\cite{wolfe}). 
This corresponds to a separation between the galaxy and the 
line-of-sight to the background quasar of less than 1 to 2$\arcs$. 
%$<$The quasar brightness at a radius of 2 -- 3$\arcs$ still exceeds 0.1\% 
%of its peak (central) brightness.$>$
Detecting a faint galaxy this close to a bright source is exceedingly difficult from the ground. 
Adaptive optics is not yet the solution (beset by poor Strehl ratios).   
Proposed interferrometric 
techniques are more suitable for detecting a compact source at a known 
position angle with respect to the quasar (and still the required 
dynamic range may be unattainable).
Ground-based imaging can discover DLA galaxies 
{\it only} at large impact parameters, greater than 25 kpc (for
H$_{o}$ = 75).  To do better, an extremely sharp and stable PSF is required, 
preferably near the diffraction limit.  This means going into space.

\begin{figure}
\plotfiddle{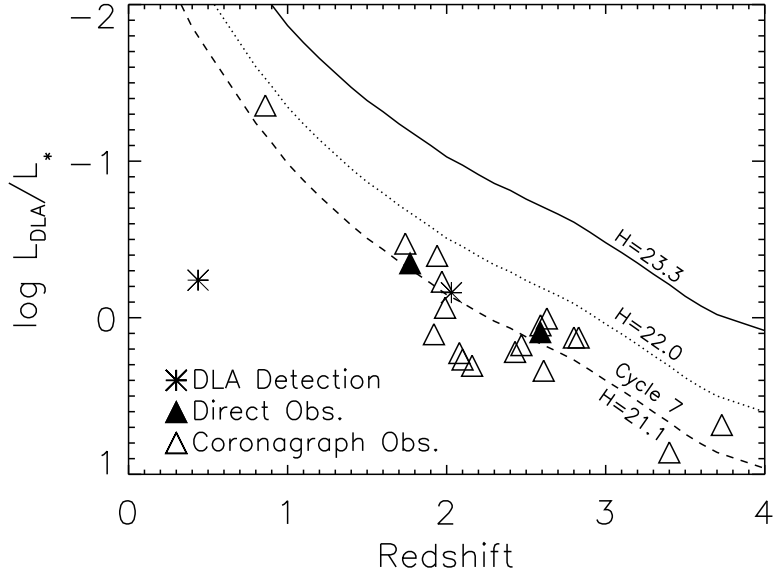}{1.9in}{90}{50}{50}{170}{-40}
\caption{Upper limits to the 1.6\mic\ continuum emission from starlight
associated with Damped \lya Absorbers, from Colbert and Malkan (\cite{com}).
Since they are all limits
(except for the two plausible detections shown by asterisks),
they are pointing upwards (towards less luminous galaxies).
These upper limits would be a factor of two tighter for any
galaxies more than 1\arcs\ away from the quasar (ie., they
understate the true sensitivity of the observations).
The dashed, dotted and solid lines illustrate typical $5-\sigma$
upper limits obtained in our Cycle 7 program, and future possible
NICMOS observations with somewhat longer, and much longer
integration times.
For twenty observed DLA systems, our HST coronographic imaging rules
out the possibility of a galaxy having typically a luminosity of
0.7--1.0 $L_*$. }
\end{figure}

We therefore used the NICMOS-2 camera on the HST in coronographic imaging mode 
to search for near-IR continuum emission from starlight associated
with two dozen DLAs.
Our results are shown in Figure 14.
Despite the approval of 95 targets, only two dozen 
quasar observations were attempted, and 
many were not as sensitive as they could have been. %
%only 12 were well executed.
This low success rate was 
because: 1) the coronograph was not fully 
operational during the first half of Cycle 7, 2) SNAPSHOT proposals utilizing the 
coronograph were not scheduled for a few months after the coronograph became 
operational, and 3) failure  
of the coronograph to autonomously acquire several of the quasars 
when they were scheduled resulted in a 
loss of those observations.  
%[possibly due to migration of the coronographic hole in response to the NICMOS dewar anomaly (cite),
%resulting in a complete loss of data. 
%This incomplete project yielded some intriguing detection upper limits, although there are 
%too few of them to have sampled the width of the luminosity function. 
Despite these difficulties, we detected two likely DLA galaxies of $\sim$ 0.7$L_\star$, and obtained upper  
limits for nearly twenty DLAs (\cite{com}).  
%The DLA near quasar LBQS 0010-0012, with z$_{abs}$=2.03, is shown in Figure 1.
%The bright field 
%star that was scaled for the PSF subtraction is to the lower left, the quasar is upper right and 
%the DLA is off at 11 o'clock at a projected  
%separation of 30 kpc.  
Our %conservative 
5$\sigma$ upper limits for the continuum fluxes of the DLAs are 
plotted in terms of $L_*$ versus redshift in Figure 14. 
Note that the vertical axis has logarithmic luminosity increasing 
downwards--higher observations are more sensitive to fainter galaxies.
The only detections of possible DLA galaxies are shown
by the two asterisks.
The lowest curve is our measured sensitivity for Cycle 7;
filled triangles represent direct imaging, while open triangles 
represent coronographic observations.  
The middle curve shows our predicted sensitivity in a 1300-second 
NICMOS integration, a half magnitude gain over our current limits.  
Finally, the upper curve is the predicted sensitivity 
of for ``limiting" 10,000-second exposures, would push detection limits 2 mags fainter.
%for four DLAs. These curves quantify how much more sensitive searches for lower redshift DLAs are. 

There were only two plausible DLA detections and the  
best upper limits were H $\sim$ 21.5 mag.
When we account for the fact that more luminous galaxies
are likely to have larger HI cross sections, we would
have expected to detect 5 or 6 of the galaxies. 
If the DLAs were drawn from the general galaxy population,
we should have detected 2 or 3 with $1.0 L_*$ or brighter,
instead of zero.
This discrepancy indicates that the luminosity function of DLAs 
is not consistent with the normal galaxy population.
The objects associated with DLAs are on average fainter
than normal galaxies we measure at either z=0 or 3.
Deeper NICMOS/HST imaging can test the possibility that
they are produced by dwarf galaxies such as the Magellanic Clouds.
%Progress in observational cosmology requires  
%the detection and study of an unbiased sample of  
%galaxy-size systems at accurately known high redshifts. 
%Such a sample is crucial to understanding the history of star formation, the galaxy  
%luminosity function, the evolution  
%of galaxies and, perhaps, galaxy formation. 
% Even if both of these techniques eventually realize their full 
%potential, they will always impose {\it strong selection biases} 
%on high-z galaxies. 
%The emission-line searches -- very dependent upon their large equivalent 
%widths -- are sensitive only to galaxies with the highest rates of 
%current star formation. Similarly, the UV-dropout technique is 
%sensitive only to blue galaxies, with large young stellar 
%populations and small internal reddenings. 
%To study the 
%objects overlooked by these biased methods requires a very different 
%{\it independent search technique}.   
%\vskip 0.1 in 
%\noindent 
     
If deeper proposed NICMOS imaging of DLA systems continues to 
set even stricter upper limits on their continuum emission,
it may be that 
DLAs trace a substantial baryon reservoir not associated with normal
galaxies.

%{\it Since DLAs are the least biased sample of galaxy-scale  
%objects at high  
%redshift, a top priority of observational cosmology should be the detection 
%and study of the starlight associated with them.} 
%%and likely hold keys to understanding the origin and evolution of galaxies.  
%To answer this, {\it the study of starlight from DLAs 
%is a top priority of observational cosmology}.  

\vskip 0.7in

I thank James Colbert, Harry Teplitz and Erin Hicks for some of the figures from our
ongoing research collaborations.


\vskip 0.7in

\begin{thebibliography}{99}
\bibitem{adel} K. Adelberger and C. Steidel 2000, ApJ 544, 218.
\bibitem{bol} Bolzonella et al. 2000, A$\&$A, 363, 476.
\bibitem{brun} R. Brunner et al. 1997 ApJLett 482, L21.
\bibitem{bunk} A. Bunker, in The Hy-Redshift Universe: 
Galaxy Formation and Evolution at High Redshift, Proc. of  conf. in 
Berkeley, CA, 21-24 June, 1999. ASP Conference Proceedings, Vol. 193, Eds.
                  by Andrew J. Bunker and Wil J. M. van Breugel (1999), p.448.
\bibitem{com} J. Colbert and M. Malkan, ApJ, in press.
\bibitem{com02} J. Colbert,  M. Malkan, and R. M. Rich ApJ, in prep.
\bibitem{con97} A. Connolly et al. 1997 ApJLett, 486, L11.
\bibitem{fern} Fernandez-Soto et al. 1999, ApJ, 513, 34.
\bibitem{fran} Francis, P. J. et al. 1996 ApJ 450, 497
\bibitem{gm} T. Glassman and M. Malkan 2002, ApJ, in press.
\bibitem{jean} Gorjian, Varoujan, Turner, Jean L., and  Beck, Sara C. 2001, ApJLett 544, L29.
\bibitem{hicks} E. Hicks, \et 2002, ApJ, submitted.
\bibitem{hogg} D.  Hogg, \et  1998, ApJ, 504, 622.
\bibitem{hop} Hopkins, \et  2000 AJ 120, 2843.
\bibitem{kaw} Kawara, \et\ 1998 A\&A 336, L9.
\bibitem{ken} R. Kennicut 1983 ApJ 272, 54.
\bibitem{koo} D. Koo 1985, AJ, 90, 418.  
\bibitem{lil} S. Lilly \et 1996, ApJ Lett 460, L1.
\bibitem{lowen} Lowenthal, J. D., Hogan, C. J., Green, R. F., Woodgate, B., Caulet, A., 
Brown, L. and Bechtold, J. 1995, ApJ 451, 484.
\bibitem{ma96} P. Madau \et 1996, MNRAS, 283, 1388. 
\bibitem{ma98} P. Madau \et 1998, ApJ, 498, 106.
\bibitem{mai} T. Maihara \et 2001, PASJ, 53, 25.
\bibitem{mat02} M. Malkan \et 2002, ApJ, in prep.
\bibitem{mat01} M. Malkan 2001, Proceedings of Fourth RESCEU International
Symposium, Our Second Look at the Immature Universe: The Infrared
View, eds. K. Sato and M. Kawasaki (Univ. Acad. Press: Tokyo), 119.
\bibitem{ms98} M.  Malkan and F. Stecker 1998, ApJ, 496, 13.
\bibitem{ms01} M.  Malkan and F. Stecker 2001, ApJ, 555, 641.
\bibitem{mtm95} M. Malkan, Teplitz, \& Mclean (MTM) 1995, ApJ, 448L, 5. 
\bibitem{mtm96} M. Malkan, Teplitz, \& Mclean 1996, ApJ, 468L, 9.
\bibitem{mann} F. Mannucci \et  1998 ApJLett, 501, L11.
\bibitem{pmc} McCarthy, P. \et\ 1999, ApJ 520, 548.
%  Rao, S.M., \& Turnshek, D.A. 1998, ApJL       500, L115
\bibitem{ouch} M. Ouchi 2001, ApJLett, 558, L83.
%  Pettini, M. {\it \et} 2001, ApJ 553, 288.
\bibitem{pet98} Pettini, M., Kellogg, M., Steidel, C.C., Dickinson,
M., Adelberger, K. L., \& Giavalisco, M., 1998, ApJ 508, 539.
\bibitem{pet} Pettini, M. {\it \et} 2001, ApJ 554, 981.
%Pettini, M., Ellison, S.L., Steidel, 
%        C.C., Shapley, A.E., \& Bowen, D.V., 2000 ApJ, 532, 65.
\bibitem{mrr} M. Rowan-Robinson 1999 ApSpSci, 266, 291.
\bibitem{ss} S. Sakai, R. C. Kennicutt, and C. Moss 2000, in Galaxy Disks and Disk
  Galaxies, eds. J. G. Funes and E. M. Corsini (ASP Conf. Ser.), 329.
\bibitem{tep98} H. Teplitz et al 1998, ApJ, 506, 519.
\bibitem{tep99} H. Teplitz,H.I, Malkan,M.A., \& McLean,I.S., 1999, ApJ 514, 33.
\bibitem{tep00} H. Teplitz et al 2000, ApJ 542, 18.
\bibitem{st96}  C. Steidel \et 1996, AJ, 112, 352.
\bibitem{st99}  C. Steidel, C.C., Adelberger, K.L., 
Giavalisco, M., Dickinson, M., \& Pettini, M. 1999, ApJ 519, 1.
\bibitem{wolfe} Wolfe, A. M. 1988, in QSO Absorption Lines:  Probing the Universe, 
        ed. J. C. Blades, D. A. Turnshek, and C. A. Norman (Cambridge: 
Cambridge University Press), 297
\bibitem{yan} Yan, L. \et\ 1999 ApJ Lett 519, 47.
\end{thebibliography}
\end{document}